\documentclass{ws-ijmpcs}

\newcommand{\dhd}{{\textstyle d}
\lower.03ex\hbox{\kern-0.38em$^{\scriptstyle-}$}\kern-0.05em{}}
\newcommand{\dbar}{{\textstyle \delta}
\lower.03ex\hbox{\kern-0.38em$^{\scriptstyle-}$}\kern-0.05em{}}
\newcommand{\half}{{1\over 2}}

\newcommand{\bark}{{\bar k}}

\newcommand{\barq}{{\bar q}}

\newcommand{\bamma}{{\bar \gamma}}

\newcommand{\calf}{{\cal F}}

\newcommand{\calu}{{\cal U}} 
\newcommand{\calv}{{\cal V}}

\newcommand{\barP}{{\bar P}}

\newcommand{\hatj}{{\hat j}}

\newcommand{\tildek}{{\tilde k}}
\newcommand{\tildeP}{{\tilde P}}  
\newcommand{\tildeq}{{\tilde q}}

\begin{document}

\markboth{Giovanni Antonio Chirilli}
{High-Energy QCD factorization from DIS to pA collisions}

\catchline{}{}{}{}{}
%
%%%%%%%%%%%%%%%%%%%%%%%%%%%%%%%%%%%%%%%%%%%%%%%%%%%%%%%%%%%%%%%%%%%%

\title{High-Energy QCD factorization from DIS to pA collisions}

\author{Giovanni Antonio Chirilli}

\address{Lawrence Berkeley National Laboratory,\\
Nuclear Science Division, \\
Berkeley, CA 94720, USA\\
gchirilli@lbl.gov}

\maketitle

%\begin{history}
%\received{Day Month Year}
%\revised{Day Month Year}
%\end{history}

\begin{abstract}
The high-energy QCD factorization for Deep Inelastic Scattering and for proton-nucleus collisions using
Wilson line formalism and factorization in rapidity is discussed. We show that in DIS the factorization in rapidity
reduces to the $k_{\rm T}$-factorization when the 2-gluon approximation is applied, provided that the composite Wilson line operator is used 
in the high-energy Operator Product Expansion. We then show that the inclusive forward cross-section in proton-nucleus collisions 
factorizes in parton distribution functions, fragmentation functions and dipole gluon distribution function at one-loop level.
\end{abstract}

\ccode{PACS numbers: 12.38.Bx, 12.38.Cy}

\section{Introduction}	

In Quantum Chromodynamics, factorization is a fundamental concept. 
It allows the applicability of perturbative methods to the calculation of scattering amplitudes 
and cross sections of hadronic processes. Well known factorization schemes are the collinear factorization and the $k_{\rm T}$-factorization.
These two factorization schemes are relevant when the hadronic matter is probed at two different limit: the Bjorken limit and the Regge 
(high-energy or small-x) limit. The collinear factorization is applied in the Bjorken limit where the dynamics of the process
is governed by incoherent interactions. On the other hand, the $k_{\rm T}$-factorization is relevant at high-energy (Regge limit) where
the dynamics of the process is dominated, instead, by coherent interactions.

\begin{figure}[tb]
\hspace{-0.3cm}
\centerline{\psfig{file=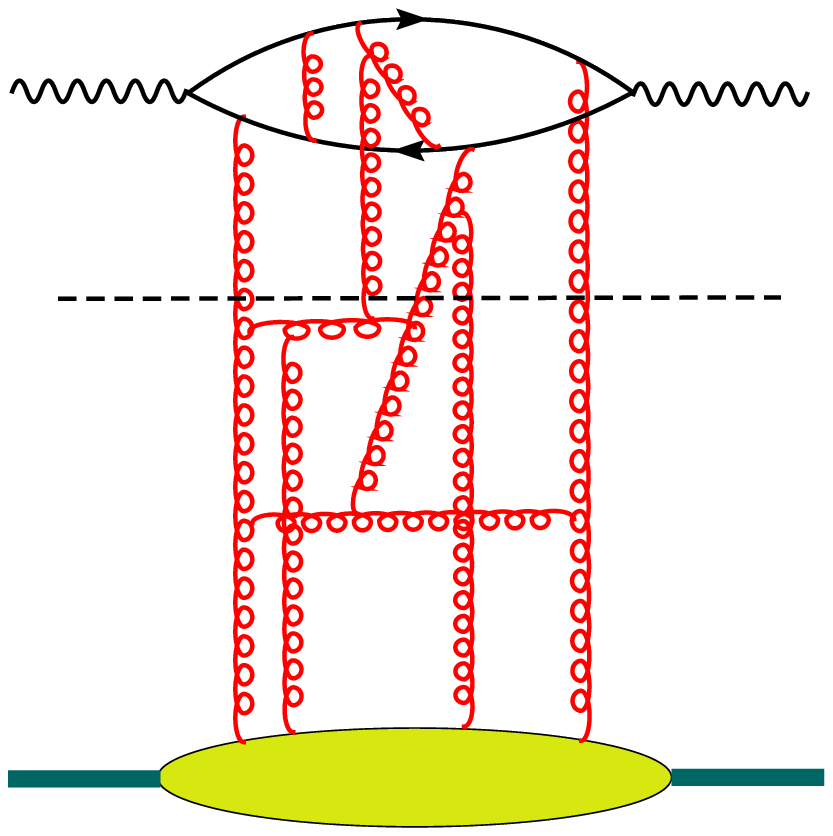,width=35mm}
\hspace{0.7cm}
\psfig{file=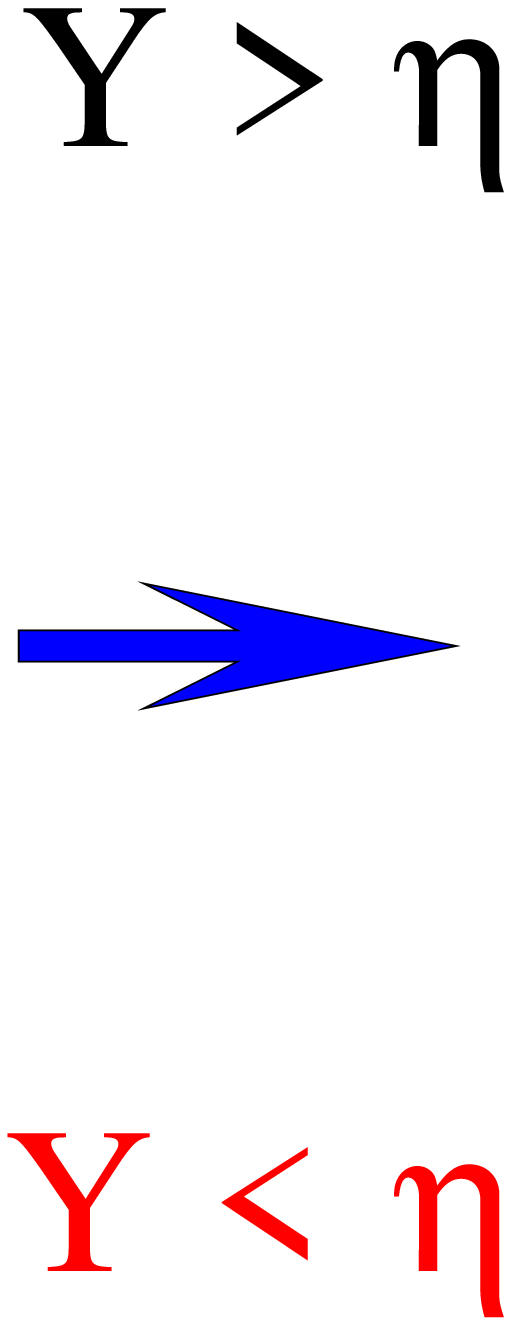,width=6mm}
\hspace{0.7cm}
\psfig{file=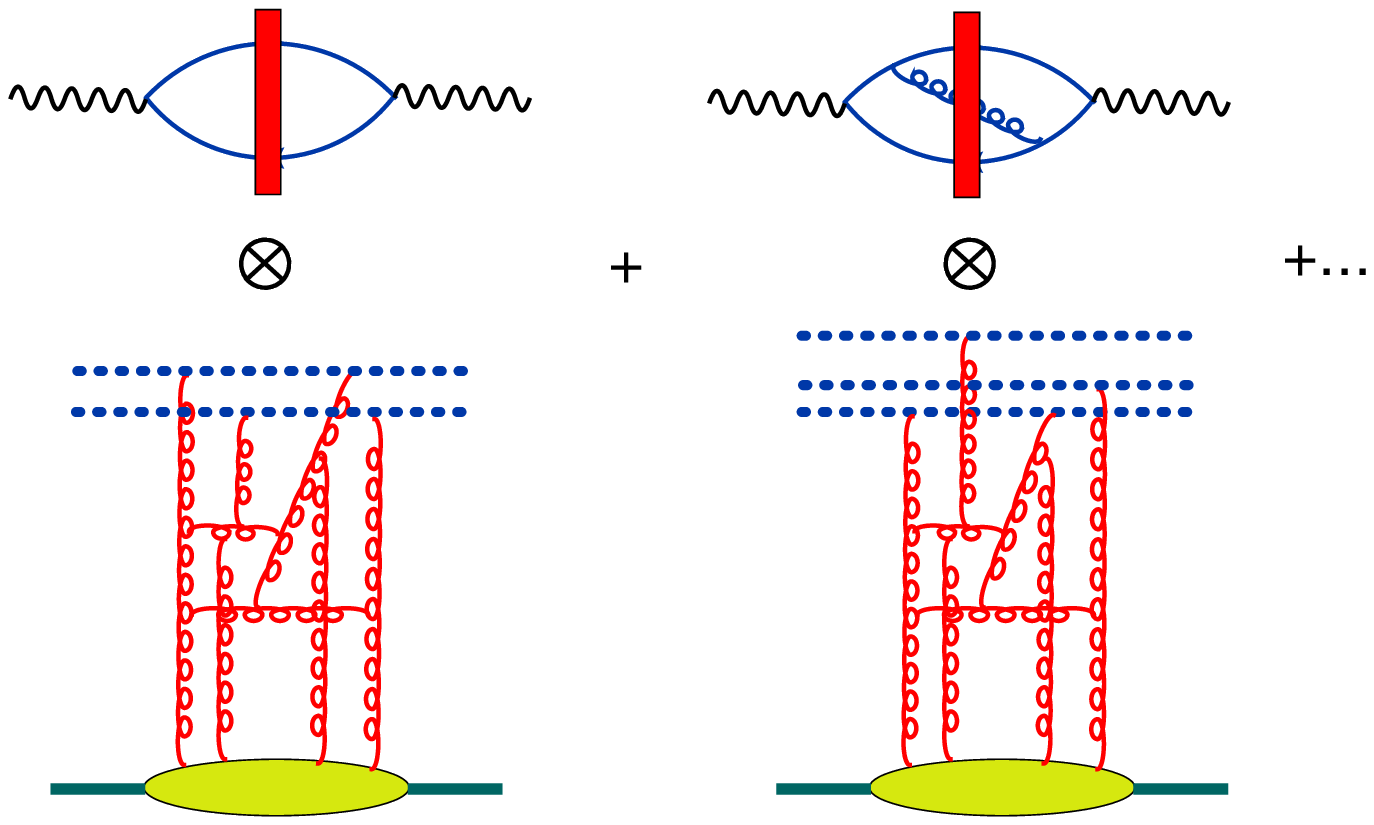,width=60mm}}
\label{OPE-expan}
\caption{Expansion of the ${\rm T}$-product of two electromagnetic currents in terms of Wilson-line operators. The blue dotted lines represent the
Wilson line operators.}
\end{figure}

At high-energy (Regge limit)
the scattering amplitude is suitably factorized in rapidity-space (see Fig. (\ref{OPE-expan})), and the 
coefficient-functions and matrix elements of non-local operators of the Operator Product Expansion at high energy, contain 
perturbative and non perturbative contributions. 
The non local operators are Wilson lines: infinite gauge link ordered along the straight line collinear to the particle's 
velocity
\begin{eqnarray}
&&\hspace{-0mm} 
U^\eta_x~=~{\rm Pexp}\Big[ig\!\int_{-\infty}^\infty\!\! du ~p_1^\mu A^\sigma_\mu(up_1+x_\perp)\Big],
\nonumber\\
&&\hspace{-0mm} 
A^\eta_\mu(x)~=~\int\!d^4 k ~\theta(e^\eta-|\alpha_k|)e^{ik\cdot x} A_\mu(k)
\label{cutoff}
\end{eqnarray}
Here $\eta$ is the particle's rapidity.
The evolution in rapidity of these operators is known as the Balitsky-equation\cite{npb96}: a non-linear evolution equation which 
generates a hierarchy of coupled equations known as the Balitsky-hierarchy\cite{npb96} (for a review see \refcite{b-review}). 
In the Color Glass Condensate formalism it 
coincides with the JIMWLK evolution equation\cite{JalilianMarian1997gr,Iancu2000hn,Ferreiro2001qy}. In the large $N_c$ limit, instead, 
the Balitsky-equation decouples and is written in a closed form that is known, in DIS case, as the Balitsky-Kovchegov (BK) 
equation\cite{npb96,yura}. 
The linear version of the BK equation is the BFKL equation\cite{bfkl}.

\section{NLO photon impact factor for DIS}

Let us now illustrate the logic of the OPE at high energies when applied to the T-product of two electromagnetic 
currents which will be relevant for DIS process. The technique we are using is the 
background field technique: the T-product of the two electromagnetic currents is considered in the background of gluon field. 
In the spectator frame the background field reduces to a shock wave (for review see \refcite{b-review}). 
In DIS, in the dipole model, the  virtual photon which  mediate the interactions
between the lepton and the nucleon (or nucleus), 
splits into a quark anti-quark pair long before the interaction with the target. The propagation of the 
quark anti-quark pair in the background of a shock wave, reduces to two Wilson lines.  
If the quark fluctuate perturbatively in a quark and a gluon before interacting with the target, then the 
number of Wilson lines increases. Formally, we can write down the expansion of the T-product of two electromagnetic currents in the following way
\begin{eqnarray}
&&\hspace{-3mm}
T\{\hat{j}_\mu(x)\hat{j}_\nu(y)\}=\int\! d^2z_1d^2z_2~I^{\rm LO}_{\mu\nu}(z_1,z_2,x,y)
[{\rm Tr}\{\hat{U}^\eta_{z_1}\hat{U}^{\dagger\eta}_{z_2}\}]^{\rm comp.}
\\
&&\hspace{-3mm}
+\int\! d^2z_1d^2z_2d^2z_3~I^{\rm NLO}_{\mu\nu}(z_1,z_2,z_3,x,y)
[{\rm tr}\{\hat{U}^\eta_{z_1}\hat{U}^{\dagger\eta}_{z_3}\}{\rm tr}\{\hat{U}^\eta_{z_3}\hat{U}^{\dagger\eta}_{z_2}\}
-N_c{\rm tr}\{\hat{U}^\eta_{z_1}\hat{U}^{\dagger\eta}_{z_2}\}] + \cdots
\nonumber
\label{OPE}
\end{eqnarray}
where $U_x = {\rm Pexp(ig\int dx^+A^-(x^+ +x_\perp)}$ is the Wilson line. 
In  Eq. (\ref{OPE}), the coefficient $I^{\rm LO}_{\mu\nu}$ represents the leading order impact factor, while the NLO impact factor
is given by the coefficient $I^{\rm NLO}_{\mu\nu}$. In QCD, Feynman diagrams at tree level are conformal invariant. 
The LO impact factor is indeed conformal invariant and it can be written in terms of conformal vectors 
$\kappa~=~{\sqrt{s}\over 2x_\ast}({p_1\over s}-x^2p_2+x_\perp)-{\sqrt{s}\over 2y_\ast}({p_1\over s}-y^2p_2+y_\perp)$ and
$\zeta_i~=~\big({p_1\over s}+z_{i\perp}^2 p_2+z_{i\perp}\big)$
\begin{eqnarray}
&&\hspace{-1cm}
\langle T\{\hat{j}_\mu(x)\hat{j}_\nu(y)\} \rangle_A~ 
=~{s^2\over 2^9\pi^6x_\ast^2y_\ast^2}\int d^2z_{1\perp}d^2z_{2\perp}
{{\rm tr}\{U_{z_1}U^\dagger_{z_2}\}\over (\kappa\cdot\zeta_1)^3(\kappa\cdot\zeta_2)^3}\times
\nonumber\\
&&\hspace{-1cm}
{\partial^2\over\partial x^\mu\partial y^\nu}
\big[2(\kappa\cdot\zeta_1)(\kappa\cdot\zeta_2)-\kappa^2(\zeta_1\cdot\zeta_2) \big]~+~O(\alpha_s)
 \label{loif}
\end{eqnarray}
Although the NLO impact factor is also made by tree level diagrams, it is not conformal invariant due 
to the rapidity divergence present at this order.
Since we regularize such 
divergence by rigid cut-off, we introduce terms which violate conformal invariance. In order to restore the symmetry we introduce 
counterterms which form the composite operator. The procedure of restoring the loss of conformal symmetry due to the regularization of the rapidity 
divergence by rigid cut-off, is analog to the procedure of restoring gauge invariance by adding counterterms to local operator when the rigid 
cut-off is used instead of dimensional regularization, that automatically preserve gauge symmetry, 
to regulate ultraviolet divergence at one loop order. 

In Eq. (\ref{OPE}), the composite operator is
\begin{eqnarray}
&&\hspace{-5mm}
[{\rm Tr}\{\hat{U}^\eta_{z_1}\hat{U}^{\dagger\eta}_{z_2}\}\big]^{\rm conf}~=~{\rm Tr}\{\hat{U}^\eta_{z_1}\hat{U}^{\dagger\eta}_{z_2}\}
\\
&&\hspace{-5mm}
+~{\alpha_s\over 4\pi}\!\int\! d^2 z_3~{z_{12}^2\over z_{13}^2z_{23}^2}
[{1\over N_c} {\rm tr}\{\hat{U}^\eta_{z_1}\hat{U}^{\dagger\eta}_{z_3}\} {\rm tr}\{\hat{U}^\eta_{z_3}\hat{U}^{\dagger\eta}_{z_2}\}
- {\rm Tr}\{\hat{U}^\eta_{z_1}\hat{U}^{\dagger\eta}_{z_2}\}]
\ln {az_{12}^2\over z_{13}^2z_{23}^2}~+~O(\alpha_s^2)
\nonumber
\end{eqnarray}
The parameter $a$ is the analog of $\mu_F$ in the usual OPE. Note also that at this order the operator proportional to the NLO impact factor
does not need to be modified. It would get a counterterm at NNLO accuracy. Using, then, the composite operator, the NLO impact factor is conformal 
invariant and it can be written entirely in terms of the conformal vectors we defined above. See Ref. \refcite{nloif} for its explicit expression. 
Such result is an analytic expression of the photon impact factor in coordinate space which is relevant
for DIS off a large nucleus where the non linear operator appearing at NLO level is relevant at high parton density 
regime\cite{Chirilli:2011zz,Beuf:2011xd}. 

\subsection{NLO photon impact factor for DIS in the $k_{\rm T}$ factorization scheme}

\begin{figure}[tb]
\hspace{-0.3cm}
\centerline{\psfig{file=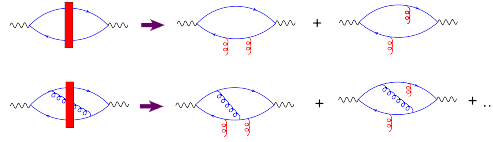,width=65mm}}
\label{2g-expan}
\caption{Diagrammatic representation of the 2-gluon approximation applied to Wilson line operators proportional 
to LO and NLO impact factor respectively.}
\end{figure}

What we are interested in is the NLO impact factor for BFKL pomeron in momentum space.
Thus, our next step, before proceeding to the calculation of the Mellin 
representation which is a useful intermediate step to get the momentum representation, 
is to obtain the linearization of result in coordinate space in the non-linear case. 
Diagrammatically, the linearization of the non linear terms of the non-linear equation is given in Fig. 
(\ref{2g-expan}). To get the right-hand-side (RHS) of Fig. (\ref{2g-expan}) we applied the 2-gluon approximation, thus, 
reproducing the diagrams of the NLO Impact factor of the usual perturbative QCD.
It turns out that the coordinate representation of the NLO impact factor
in the linearized case 
can be written as a linear combination of five conformal tensor structures\cite{nloif,Cornalba:2009ax}. 

%In Ref. \refcite{Cornalba:2009ax} it was indeed predicted that any impact factor can be written as a linear combination of the 
%same conformal tensor structures.
The projection of the impact factor on the Lipatov eigenfunctions with conformal spin $0$ is related to
the unpolarized structure function for DIS. While the projection on the Lipatov eigenfunction with conformal spin $2$ is related to the polarized 
structure function. The result of the Mellin representation can be found in Ref. \refcite{nloifms}.
Once we have performed the Mellin representation, we are ready to perform the Fourier transform in momentum space. 
The result of the Fourier transform is\cite{nloifms}
\begin{eqnarray}
&&\hspace{-1cm}
I^{\mu\nu}(q,k_\perp)
={N_c\over 32}\!\int\! {d\nu\over \pi\nu}{\sinh\pi\nu\over (1+\nu^2)\cosh^2\pi\nu}
\Big({k_\perp^2\over Q^2}\Big)^{\half-i\nu}
\times
\nonumber\\
&&\hspace{-1cm}
\Big\{
\Big[\Big({9\over 4}+\nu^2\Big)\Big(1+{\alpha_s\over\pi}+{\alpha_sN_c\over 2\pi}\calf_1(\nu)\Big)P_1^{\mu\nu}
\nonumber\\
&&\hspace{-1cm}
+\Big({11\over 4}+3\nu^2\Big)\Big(1+{\alpha_s\over\pi}+{\alpha_sN_c\over 2\pi}\calf_2(\nu)\Big)P_2^{\mu\nu}
\Big]
\nonumber\\
&&\hspace{-1cm}
+
{{1\over 4}+\nu^2\over 2k_\perp^2}
\Big(1+{\alpha_s\over\pi}+{\alpha_sN_c\over 2\pi}\calf_3(\nu)\Big)\big[\tildeP^{\mu\nu}\bark^2+\barP^{\mu\nu}\tildek^2\big]\Big\}
\label{nloifmome}
\end{eqnarray}
where
\begin{eqnarray}
&&\hspace{-1mm}
P_1^{\mu\nu}~=~g^{\mu\nu}-{q_\mu q_\nu\over q^2};
~~~~~
P_2^{\mu\nu}~=~{1\over q^2}\Big(q^\mu-{p_2^\mu q^2\over q\cdot p_2}\Big)\Big(q^\nu-{p_2^\nu q^2\over q\cdot p_2}\Big)
\nonumber\\
&&\hspace{-1mm}
\barP^{\mu\nu}~=~\big(g^{\mu 1}-ig^{\mu 2}-p_2^\mu{\barq\over q\cdot p_2}\big)\big(g^{\nu 1}-ig^{\nu 2}-p_2^\nu{\barq\over q\cdot p_2}\big)
\nonumber\\
&&\hspace{-1mm}
\tildeP^{\mu\nu}~=~\big(g^{\mu 1}+ig^{\mu 2}-p_2^\mu{\tildeq\over q\cdot p_2}\big)\big(g^{\nu 1}+ig^{\nu 2}-p_2^\nu{\tildeq\over q\cdot p_2}\big)
\nonumber
\end{eqnarray}
and
\begin{eqnarray}
&&\hspace{-0mm}
\calf_{1(2)}(\nu)~=~\Phi_{1(2)}(\nu)+\chi_\gamma\Psi(\nu),
~~~~
\calf_3(\nu)~=~F_6(\nu)+\Big(\chi_\gamma-{1\over\bamma\gamma}\Big)\Psi(\nu),
\nonumber\\
&&\hspace{0mm}
\Psi(\nu)~\equiv~\psi(\bamma)+2\psi(2-\gamma)-2\psi(4-2\gamma)-\psi(2+\gamma),
\nonumber\\
&&\hspace{-0mm}
F_6(\gamma)~=~F(\gamma)-{2C\over \bamma\gamma}-1-{2\over\gamma^2}-{2\over\bamma^2}
-3{1+\chi_\gamma-{1\over\gamma\bamma}\over 2+\bamma\gamma},
\nonumber\\
&&\hspace{-1mm}
\Phi_1(\nu)~=~F(\gamma)+{3\chi_\gamma\over 2+\bamma\gamma}+1+{25\over 18(2-\gamma)}
+~{1\over 2\bamma}-{1\over 2\gamma}-{7\over 18(1+\gamma)}
+{10\over 3(1+\gamma)^2}
\nonumber\\
&&\hspace{-1mm}
\Phi_2(\nu)~=~F(\gamma)+{3\chi_\gamma\over 2+\bamma\gamma}+1+{1\over 2\bamma\gamma}
-~{7\over 2(2+3\bamma\gamma)}
+{\chi_\gamma\over 1+\gamma}+{\chi_\gamma(1+3\gamma)\over 2+3\bamma\gamma},
\nonumber\\
&&\hspace{-1mm}
F(\gamma)~=~{2\pi^2\over 3}-{2\pi^2\over\sin^2\pi\gamma}-2C\chi_\gamma+{\chi_\gamma -2\over\bamma\gamma}
\nonumber
\end{eqnarray}

In order to obtain the full expression in momentum space of the NLO DIS amplitude, we need to perform also the Fourier transform of the NLO 
linearized BK equation  for the dipole form of the unitegrated gluon distribution $\calv(z)= z^{-2}\calu(z)$ where 
$\calu(x,y)= 1-N_c^{-1}{\rm tr}\{U(x_\perp U^\dagger(y_\perp)\}$.  The $k_{\rm T}$ factorized formula of the 
NLO amplitude for DIS is\cite{nloifms}
\begin{eqnarray}
&&\hspace{-0mm}
\int\! d^4x ~e^{iqx} \langle p|T\{\hatj_{\mu}(x)\hatj_\nu(0)\}|p\rangle~
=~{s\over 2}\int\! {d^2k_\perp\over k_\perp^2}~ I_{\mu\nu}(q,k_\perp) \calv_{a_m=x_B}(k_\perp)
\label{ktfacv}
\end{eqnarray}
where $I_{\mu\nu}(q,k_\perp)$ is given in Eq. (\ref{nloifmome}) and the evolution of the operator $\calv_{a_m=x_B}(k_\perp)$ can be found in Ref. 
\refcite{nloifms} and it is in agreement with
result of Ref. \refcite{nlobfkl,nlobk} (see also Ref. \refcite{nlobksym,nlobfklconf,nloampN4}).
The $k_{\rm T}$ factorized formula (\ref{ktfacv}) is a consequence of the factorization in rapidity 
of the DIS amplitude obtained through the high-energy OPE with composite Wilson line operator in the 2-gluon approximation.
In other words, the factorization in rapidity of the DIS amplitude naturally reduces to the $k_{\rm T}$-factorization 
when the 2-gluon approximation is applied.

\section{One-loop factorization for inclusive hadron production in pA collision\label{section3}}

In proton-nucleus collision in the forward region the two factorization schemes, the collinear factorization and the rapidity factorization,
enter on the same footing: the proton is treated as a diluted system which,
using collinear factorization, emits a quark or a gluon that eventually scatters off a dense target like a large nucleus.
At this point it is important to know whether
the naive LO factorization formula of the cross-section written as a convolution of the parton distributions and fragmentation functions and
dipole gluon distribution holds also at NLO level. 

\begin{figure}[tb]
\begin{center}
\includegraphics[width=44mm]{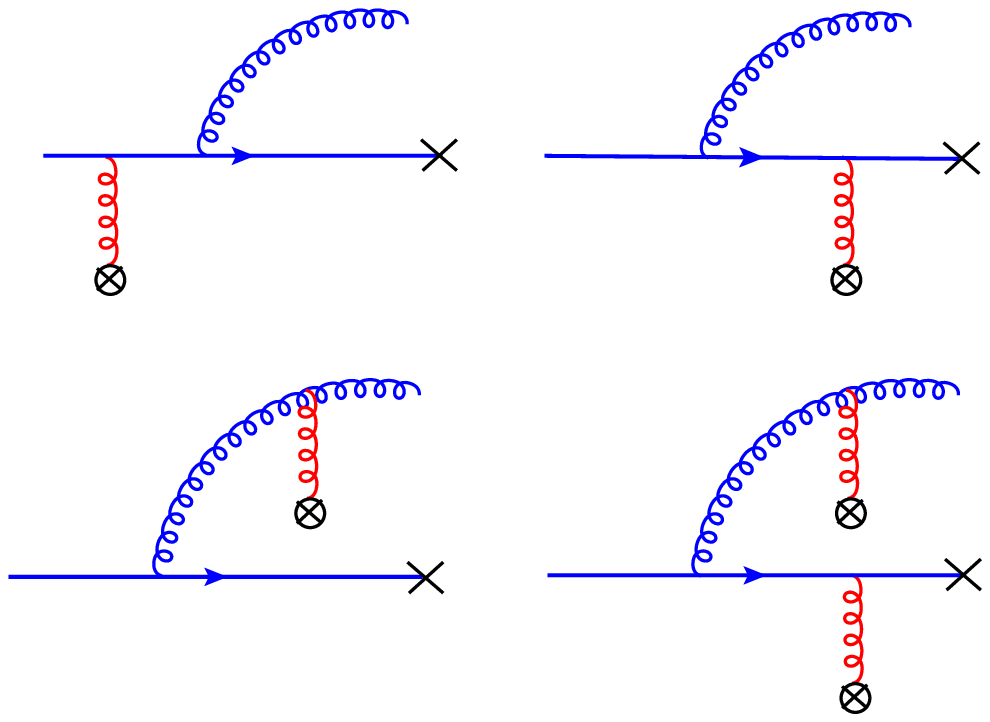}
\hspace{0.8cm}
\includegraphics[width=34mm]{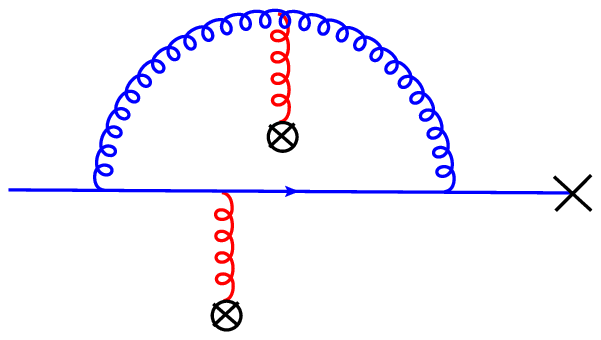}
\end{center}
\caption{Real and virtual Feynman diagrams for quark-channel at one-loop order.}
\label{one-loop-incl}
\end{figure}

To illustrate the one loop calculation, let us consider the quark-channel. 
The result of the calculation of the real and virtual Feynman diagrams at one loop order (shown in Fig. (\ref{one-loop-incl})) is 

\begin{eqnarray}
&&\hspace{-0.7cm}-\frac{\alpha_{s}N_{c}}{2\pi ^{2}}\int_0^1 \frac{d\xi^{\prime}}{1-\xi^{\prime}}%
\int \frac{d^{2}b_{\perp }}{(2\pi )^{2}}e^{-ik_{\perp }\cdot r_\perp}\frac{(x-y)_\perp^2}{(x-b)^2_\perp(y-b)^2_\perp} 
\left[S^{(2)}(x_{\perp },y_{\perp })-S^{(4)}(x_{\perp },b_{\perp},y_{\perp })\right]  \nonumber \\
&&\hspace{-0.7cm} + \frac{\alpha_sC_F}{2\pi}\int_{\tau/z}^1 d\xi \left(-\frac{1}{\epsilon}\right) \left[\mathcal{P}_{qq}(\xi) 
e^{-ik_\perp\cdot r_\perp}+\mathcal{P}_{qq}(\xi)\frac{1}{\xi^2}e^{-i\frac{k_\perp}{\xi}\cdot 
r_\perp} \right]\frac{1}{(2\pi )^{2}}S^{(2)}(x_{\perp },y_{\perp}) 
\label{1-loop}
\end{eqnarray}
where
\begin{eqnarray}
&&S^{(2)}(x_\perp,y_\perp)=\frac{1}{N_c}\langle{U}(x_\perp){U}^\dagger(y_\perp)\rangle_Y\nonumber\\
&&S^{(4)}(x_\perp,b_\perp,y_\perp)=\frac{1}{N_c^2}\langle \mathrm{Tr}[{U}(x_\perp)
{U}^\dagger(b_\perp)] \mathrm{Tr}[{U}(b_\perp){U}^\dagger(y_\perp)]\rangle_Y
\nonumber
\end{eqnarray}
In Eq. (\ref{1-loop}) we recognize the collinear divergences associated to the parton distributions and fragmentation functions, and the 
rapidity divergence associated the dipole gluon distribution.
Thus, the collinear divergence is absorbed into the renormalization of the quark distribution $[q(x,\mu)]^{1-{\rm loop}}$
and fragmentation functions $[D_{h/q}(z,\mu)]^{1-{\rm loop}}$ reproducing the DGLAP evolution equation for each of the distribution function.
While the rapidity divergence can be absorbed into the renormalization of the dipole gluon distribution
$[S^{(2)}(x_\perp,y_\perp)]^{1-{\rm loop}}$ reproducing in this way the Balitsky-BK evolution equation. 
The factorized formula for the one-loop cross section is
\begin{eqnarray}
&&\hspace{-5mm}\frac{d^3\sigma ^{p+A\to h+X}}{dyd^2p_\perp}= \int \frac{dz}{z^2}\frac{dx}{x}\,\xi\, 
x \,[q(x,\mu)]^{1-{\rm loop}} [D_{h/q}(z,\mu)]^{1-{\rm loop}}\int \frac{d^2x_\perp
d^2y_\perp}{\left(2\pi\right)^2}\times \nonumber\\
&&\hspace{-5mm}\Bigg\{[S^{(2)}(x_\perp,y_\perp)]^{1-{\rm loop}}
\left[ \mathcal{H}_{2qq}^{(0)}
+\frac{\alpha_s}{2\pi}\mathcal{H}_{2qq}^{(1)}\right]
+
\int \frac{d^2b_\perp}{(2\pi)^2}
S^{(4)}(x_\perp,b_\perp,y_\perp)\frac{\alpha_s}{2\pi}
\mathcal{H}_{4qq}^{(1)}\Bigg\}
\label{1loop-cross}
\end{eqnarray}

Equation (\ref{1loop-cross}) represents the QCD factorization for hard processes in the saturation formalism at 
one loop level\cite{Chirilli:2011km,Chirilli:2012jd}. The explicit expression for the hard coefficients can be found in Ref. 
\refcite{Chirilli:2011km,Chirilli:2012jd}.
The factorization formula for inclusive hadron production in proton-nucleus collisions is diagrammatically represented 
in Fig. (\ref{one-loop-incl}).

It is plausible to think that the QCD factorization formula for inclusive hadron production in pA collision 
obtained at one-loop level holds at any order. In this case we can formally write down the expression for the NNLO cross-section as

\begin{eqnarray}
&&\hspace{-1cm}\frac{d^3\sigma ^{p+A\to h+X}}{dyd^2p_\perp}= \int \frac{dz}{z^2}\frac{dx}{x}\,\xi\, 
x \,[q(x,\mu)]^{2-{\rm loop}} [D_{h/q}(z,\mu)]^{2-{\rm loop}}\int \frac{d^2x_\perp
d^2y_\perp}{\left(2\pi\right)^2}\times 
\nonumber\\
&&\hspace{-1cm}\Bigg\{[S^{(2)}(x_\perp,y_\perp)]^{2-{\rm loop}}
\left[ \mathcal{H}_{2qq}^{(0)}
+\frac{\alpha_s}{2\pi}\mathcal{H}_{2qq}^{(1)}+\frac{\alpha_s^2}{(2\pi)^2}\mathcal{H}_{2qq}^{(2)}\right]
\nonumber\\
&&\hspace{-1cm}
+
\int \frac{d^2b_\perp}{(2\pi)^2}
[S^{(4)}(x_\perp,b_\perp,y_\perp)]^{1-{\rm loop}}
\left[\frac{\alpha_s}{2\pi}\mathcal{H}_{4qq}^{(1)}
+\frac{\alpha_s^2}{(2\pi)^2}\mathcal{H}_{4qq}^{(2)}
\right]
\nonumber\\
&&\hspace{-1cm}
+\int {d^2b\over (2\pi)^2} {d^2\omega\over (2\pi)^2} 
S^{(6)}(x_\perp, b_\perp, \omega_\perp, y_\perp)
\frac{\alpha_s^2}{(2\pi)^2}\mathcal{H}_{6qq}^{(2)}
\Bigg\}\nonumber
\end{eqnarray}

We note that at NNLO the parton distributions and fragmentation functions follow the NLO DGLAP evolution equation. While the 
dipole gluon distributions follow the Balitsky-JIMWLK evolution equation. In particular we notice that at NNLO the $S^{(2)}$
distribution follow the NLO B-JIMWLK, $S^{(4)}$ follow the LO B-JIMWLK, while at this order a new operator, $S^{(6)}$ 
(six-Wilson line operators with arbitrary white arrangements of color indices), appears
and has no evolution. The structure of the $S^{(n)}$ operators is a hierarchy of evolution equations \textit{i.e.} the Balitsky-hierarchy.

\section{Conclusions}

\begin{figure}[tb]
\begin{center}
\hspace{-0.4cm}
\includegraphics[width=58mm]{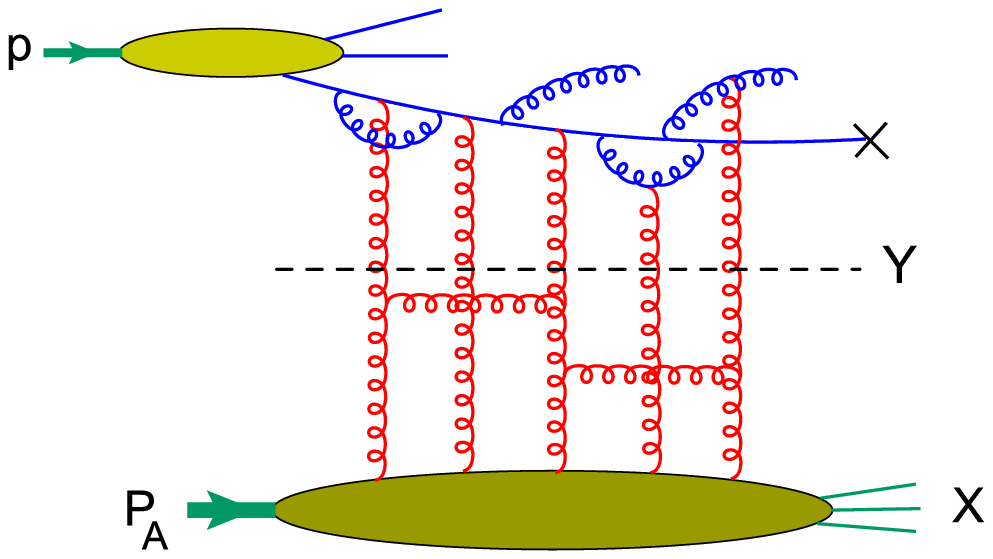}
\hspace{0.3cm}
%\begin{center}
\includegraphics[width=5mm]{freccia1a}
%\end{center}
\hspace{0.3cm}
\includegraphics[width=53mm]{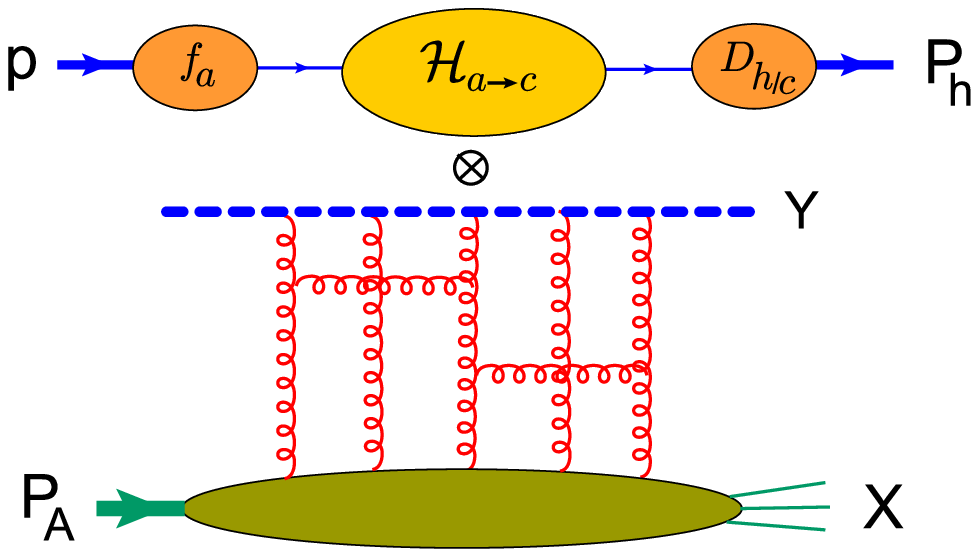}
\end{center}
\caption{QCD factorized cross section for inclusive hadron production in pA collisions. On the left-hand-side of the picture 
the black dashed line represents factorization in rapidity at point $Y$. On the right-hand-side of the picture
the dashed blue line represents a Wilson line at rapidity $Y$.}
\end{figure}

The main results we have presented here are the NLO $k_{\rm T}$-factorization formula for DIS 
and the QCD factorized formula for inclusive hadron production in pA collision.
We noticed that the $k_{\rm T}$ factorization formula for DIS, Eq. (\ref{ktfacv}), is a natural consequence of the rapidity factorization in the 
2-gluon approximation within the OPE at high energy with composite Wilson line operator. 
Result (\ref{nloifmome}) is an analytic expression of the NLO impact factor in momentum space. Before, the NLO impact factor was available only as
a combination of numerical and analytical expressions\cite{bart1}.

In section (\ref{section3}) we have presented the QCD factorization result for the inclusive hadron production in pA collisions. 
We have shown that the cross section can be written in a factorized form where the parton distributions and fragmentation 
functions follow the DGLAP evolution equation, while the operators $S^{(n)}$, due to the multiple interactions of the emitted partons with 
the target, evolve with the B-JIMWLK evolution equation. At the end we have also presented a formal expression of what would look like the 
cross section for inclusive hadron production in pA collision at two-loop level.

The author is grateful to the organizers of the QCD evolution 2012 workshop for warm hospitality and financial support.

\end{document}